\documentclass[12pt]{article}

\usepackage{amssymb,amsfonts,amstext,graphics,graphicx,subfigure}
\usepackage{setspace}
\usepackage{mathtools}

\usepackage[colorlinks]{hyperref}
\usepackage{color}

\usepackage{mathbbol}

\def\XXint#1#2#3{{\setbox0=\hbox{$#1{#2#3}{\int}$ }
\vcenter{\hbox{$#2#3$ }}\kern-.5\wd0}}

\def\calf{\mathcal{F}}

\def\calh{\mathcal{H}}

\def\cals{\mathcal{S}}

\def\bq{\begin{equation}}
\def\eq{\end{equation}}
\def\bqy{\begin{eqnarray}}
\def\eqy{\end{eqnarray}}

\def\bal#1\eal{\begin{align}#1\end{align}}

\def\de{\delta}

\def\ep{\epsilon}

\def\et{\eta}

\def\ka{\kappa}
\def\la{\lambda}
\def\La{\Lambda}

\def\si{\sigma}

\def\ze{\zeta}

\def\p{\partial}

\def\nn{\nonumber}

 \usepackage[margin=1.2in]{geometry}

 \usepackage{yfonts}
\usepackage[letterspace=45]{microtype}

\begin{document}

\pagestyle{empty}

\noindent This document is a re-issue in Latex of an original typed manuscript issued as a report:

\begin{quote}

P. J. Morrison, “Some Observations Regarding Brackets and Dissipation,” Center for Pure and Applied Mathematics Report PAM–228, University of California, Berkeley (1984).

\end{quote}

\noindent Care was taken to replicate the original.   Mistakes in the original are in an Errata  added at the end of this document.

\clearpage
\begin{centering}
\begin{figure}[htb]
\hspace{-.35in}\includegraphics[scale=.8]{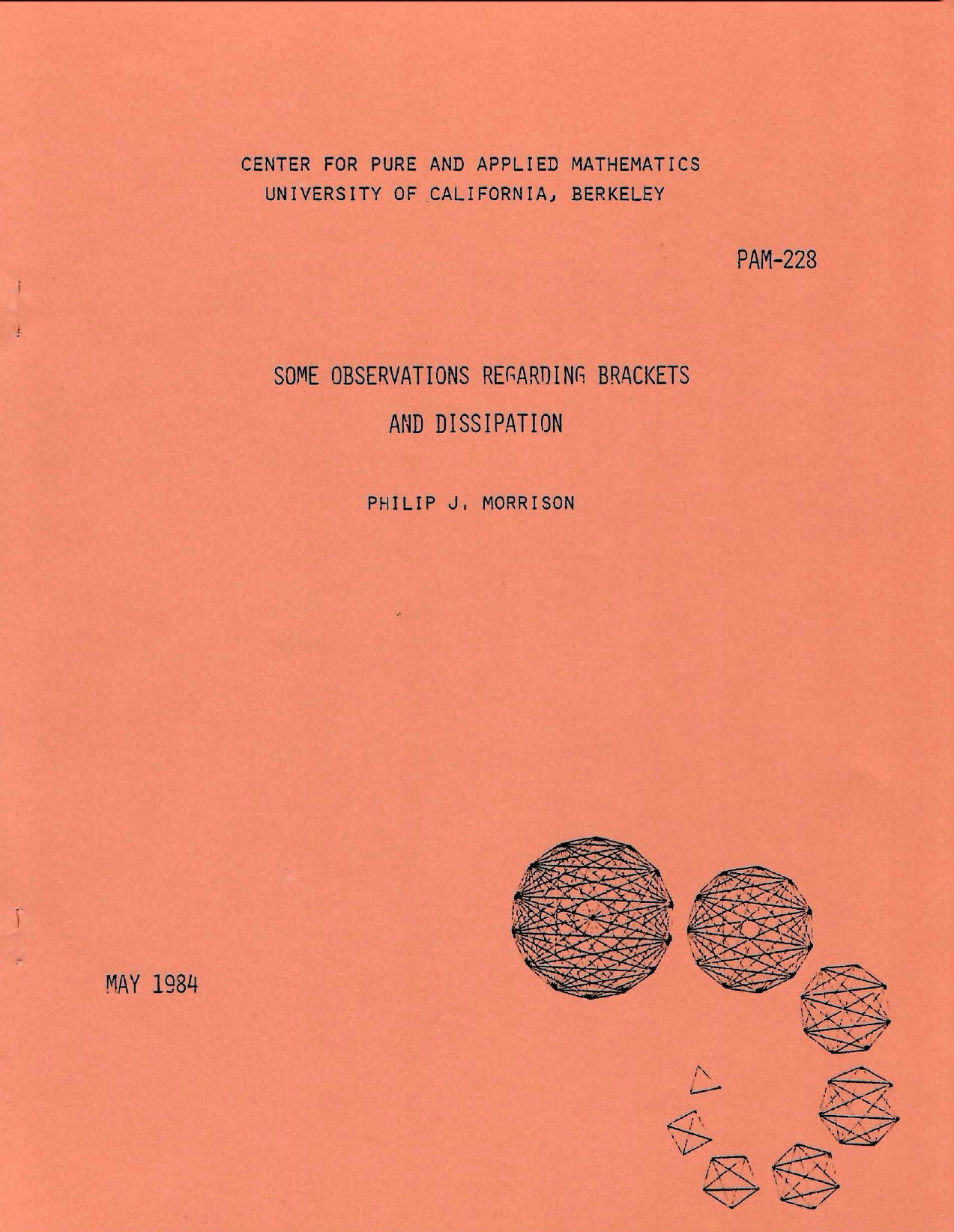}
\label{fig:tba}
\end{figure}
\end{centering}

\clearpage
\begin{centering}
\begin{figure}[htb]
\hspace{-.35in}\includegraphics[scale=.8]{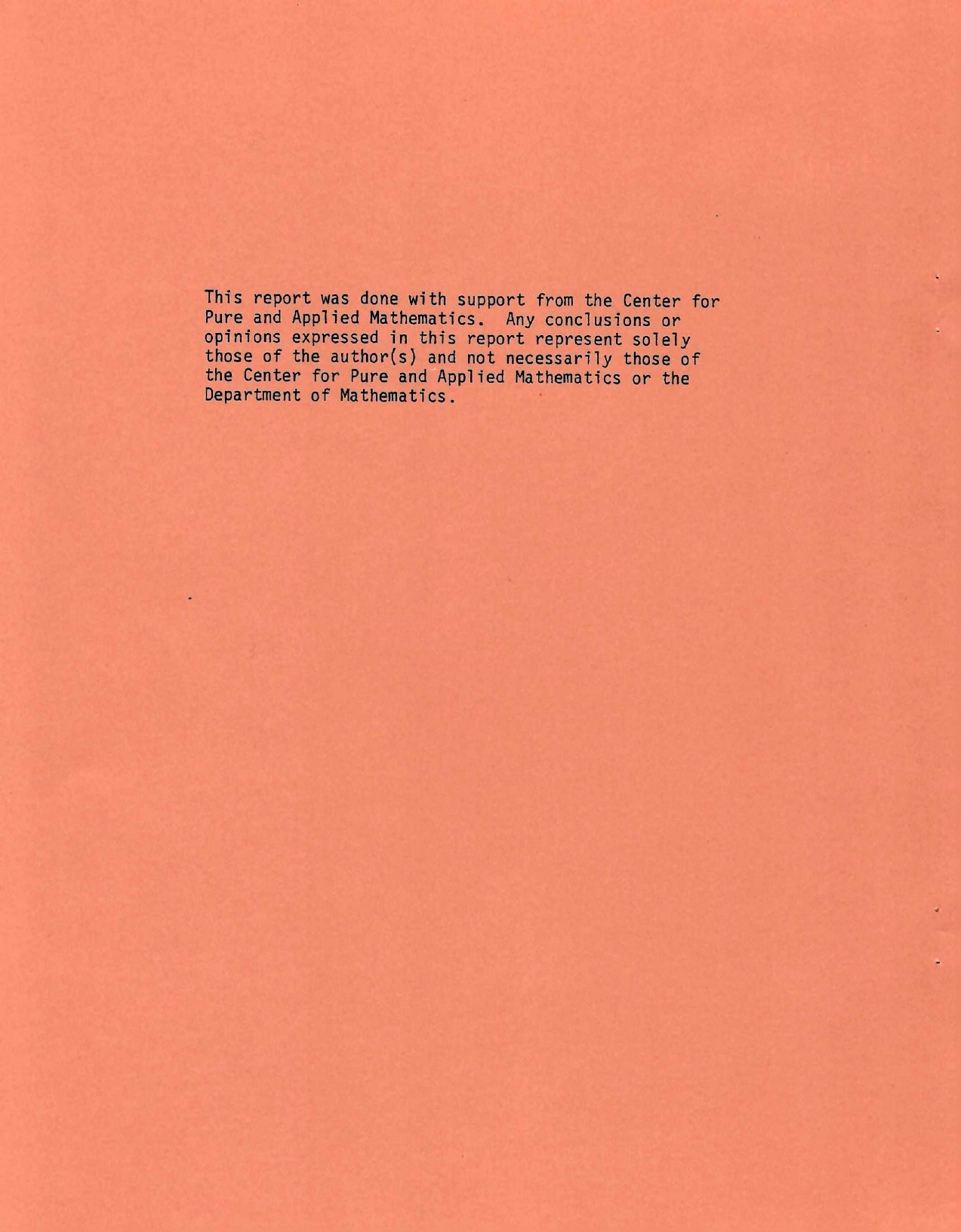}
\label{fig:tba2}
\end{figure}
\end{centering}
\clearpage

\textls{

\centerline{SOME OBSERVATIONS REGARDING BRACKETS}
\centerline{AND DISSIPATION}

\vspace*{.75 in}

 \centerline{Philip J. Morrison$^+$}

  \centerline{Department of Mathematics}
  
  \centerline{University of California}
  
  \centerline{Berkeley, CA 94720}
}

 \vspace*{.75 in}
 
 \doublespace

\noindent \underline{Abstract}

\hspace{.4in}  \textls{Some ideas relating to a bracket formulation for dissipative systems

are considered. The formulation involves a bracket that is analogous to a

generalized Poisson bracket, but possesses a symmetric component. Such a 

bracket is presented for the Navier-Stokes equations.}

\vspace{3 in}

\singlespace
\noindent${}^+$Permanent address: Department of Physics and Institute for Fusion Studies, 

\hspace{1.25 in} University of Texas at Austin, Austin, TX, 78712.

 \clearpage
 
   \setstretch{1.9}
\doublespace

 \begin{flushleft}

  \textls{
  
 \vspace*{-.34 in}

\hspace{.4in}Many of the fundamental nondissipative equations describing fluids \\
and plasmas have been shown to be Hamiltonian field  theories in terms of generalized Poisson brackets (GPB). For review see [1-4]. Here we discuss \\
a formalism for entropy producing conservative systems. As an example, the Navier-Stokes equations are considered. (This report is a companion to [5] where plasma kinetic equations are treated. A nonconservative system was discussed in [6]. Other formalisms were presented in [7-10].)
 
\hspace{.4in}Recall that a GPB is a bilinear, antisymmetric operator that is a
derivation on functionals and satisfies the Jacobi identity.
The GPB
\\
 need not be the usual Poison bracket; hence fields that do not possess standard or canonical form can sometimes still be expressed as follows:
 \medskip
\[
\frac{\p \psi^i}{\p t}= \{\psi^i, \calh\}\qquad i= 1,2,\dots N\qquad\,, \eqno(1)
\]
 \smallskip
where the Hamiltonian functional $\calh$  is the ``generator of time translation" \\
and the quantities $\psi^i$ are the field components. For two  functionals $F$ \\
and $G$ \, GPB's typically  have the form
 \medskip
\[
\{F,G\}=\int \frac{\de F}{\de \psi^i} \, O^{ij}\, \frac{\de G}{\de \psi^j} \,  d\tau \quad,\eqno (2)
\]
 \smallskip
where ${\de F}/{\de \psi^i}$, the functional derivative, is defined by $\left.\frac{d}{d\ep} \ F[\psi^i +\ep \de \psi]\right|_{\ep=0}= \int \frac{\de F}{\de \psi^i}\ \de\psi\, \, d\tau$. $d\tau$  is a volume element; and  $O^{ij}$\  is an  \\
operator that in light of antisymmetry must be anti-self-adjoint. 

\hspace{.4in}Systems that are dissipative would not a priori be expected to fit\\ into the form of Eq.~(1). Indeed it is not clear what functional should

\clearpage

\vspace*{-1.0 in}

\centerline{-2-}

\vspace*{.65 in}

be the ``generator of time translation", and which algebraic properties of a binary bracket operator will lead to a rich structure. 

\hspace{.4in} We address the first point above by recalling that in classical thermo-
\\
dynamics the equilibrium state can be obtained by either the energy or
\\
entropy extremum principles. In this sense we view the energy, a function
\\
of the extensive variables, as the ``generator of equilibria", or alternatively\\
the entropy can generate equilibria. Moreover, additional extremum principles exist in terms of the thermodynamic potentials. For dynamical systems an
extension of this is to choose from among these quantities the ``generator\\
 of time translation".

\hspace{.4in}In particular an appealing choice is a quantity we call the "generalized free energy". In the energy formulation of thermodynamics the equilibrium state is obtained by extremizing the energy at constant entropy. This can
\\
be achieved by varying the following:
\medskip
\[
F_\la = E + \la S \qquad, \eqno(3)
\]
\medskip
where $E$ is the energy, $S$ is the entropy and $\la$ is a Lagrange multiplier.
\\  A natural generalization of this for dynamical systems is to add to the
\\
Hamiltonian quantities known as Casimirs or ``generalized entropy"  functionals. 
\\
These are functionals that, due to degeneracy in a GPB, are conserved for\\
all Hamiltonians; i.e.\  they commute with all functionals. Such quantities,
\\
 independent of the GPB formalism, have previously been used to obtain varia-
 \\
 tional principles for plasma equilibria [11-14]; such principles are useful for
 \\
  obtaining linear stability criteria. Recently, using the GPB formalism, nonlinear
  
  \clearpage

\vspace*{-1.0 in}

\centerline{-3-}

\vspace*{.65 in}

stability results have been obtained using Casimirs [15,16]. Thus\\
generalizing Eq.\ (3) we obtain
\medskip
\[
\calf =\calh +\cals\eqno(4)
\]
\medskip
where $\cals$ is a Casimir.  (Observe that we have dropped the Lagrange
\\
 multiplier since  typically Casimirs involve free functions; see Eq. (22)
 \\
 below.)  The reason the quantity $\calf$ of Eq. (4) is an appealing 
 \\
 ``generator of time translation" is that by analogy \underline{critical points of $\calf$}\\
 \underline{correspond to both thermodynamic and dynamic equilibria}. $\calf$ so defined\\
 is what we have termed the ``generalized free energy."

\hspace{.4in}It remains to describe the binary bracket operator that together with\\ 
$\calf$ produces the equations of motion; i.e., in the form\\
\medskip
\[
\frac{\p\psi^i}{\p t} =\{\{\psi^i,\calf\}\}\qquad, \eqno(5)
\]
\medskip
where the double braces are used for the dissipative generalization of \\
Eq.  (2). Just as any operator can be split into self-adjoint and anti-
\\
self-adjoint parts, we split the bracket of Eq. (5) into the sum of an
\\
 antisymmetric GPB and a symmetric component. For two functionals $F$\\
  and $G$ we have
  \medskip
  \[
\{\{F,G\}\} =\{F,G\} +(F,G)\eqno(6)
\]
\medskip
where $\{F,G\}$ has the form of Eq. (2) with an anti-self-adjoint operator

 \clearpage

\vspace*{-1.0 in}

\centerline{-4-}

\vspace*{.50 in}

  $O^{ij}$ and $(F,G)$ is given by
  \medskip
 \[
 (F,G) =\int \frac{\de F}{\de \psi^i} M^{ij}   \frac{\de G}{\de \psi^j} \ d\tau \qquad. \eqno(7)
 \]
 \medskip
  Here, $M^{ij}$  is to be self-adjoint and hence $(F,G)$ is symmetric under the\\
   interchange of $F$ and $G$.
   
\hspace{.4in}Equation (5) thus becomes
\medskip
\[
\frac{\p\psi^i}{\p t} =\{\psi^i,\calf\} + (\psi^i, \calf) = (O^{ij} + M^{ij}) \frac{\de \calf}{\de \psi^j} 
\qquad, \eqno(6)
\]
\medskip
From Eq. (6) it is clear that critical points of $\calf$; i.e. points where\\
${\de \calf}/{\de \psi^i}= 0$, correspond to dynamical equilibrium, since clearly $\p \psi^i/\p t= 0$. \\
 Also Eq. (6) can be rewritten as
 \medskip
\[
\frac{\p\psi^i}{\p t} =\{\psi^i,\calh\} + (\psi^i, \calf) 
\qquad, \eqno(7)
\]
\medskip 
since the difference between $\calh$ and $\calf$ is a Casimir. From Eq. (7) we
\\
see that the dynamics is split into Hamiltonian and non-Hamiltonian parts. \\
Moreover, if the symmetric bracket has the degeneracy property $(\calh,G) = 0$\\
 for all functionals $G$, Eq. (7) becomes
 \medskip
 \[
\frac{\p\psi^i}{\p t} =\{\psi^i,\calh\} + (\psi^i, \cals) 
\qquad. \eqno(8)
\]
\medskip  
Thus the time rate of change of the generalized entropy is given by
 \[
\frac{d\cals }{d t} =  (\cals, \cals) 
\qquad. \eqno(9)
\]
 
 \clearpage

\vspace*{-1.0 in}

\centerline{-5-}

\vspace*{.65 in}

 From Eq. (9) it is clear that definiteness of the symmetric bracket is \\
  equivalent to an H-theorem. The ideas of degeneracy and definiteness\\ 
first appeared in [7] and were subsequently employed in [5, 8-10].

\hspace{.4in}We now consider the Navier-Stokes equations
 \medskip
\bal
&\frac{\p v_i}{\p t} =-v_k\frac{\p v_i}{\p x_k} -\frac1{\rho} \frac{\p p}{\p x_i} 
+ \frac1{\rho} \frac{\p \si_{ik}}{\p x_k}   \tag{10}  
\\[4.5ex]
&\frac{\p s}{\p t} =  -v_k \frac{\p s}{\p x_k} + \frac{\si_{ik}}{\rho T} \frac{\p v_i}{\p x_k} 
- \frac1{\rho T} \frac{\p q_k}{\p x_k}
\tag{11} 
\\[4.50ex]
&\frac{\p \rho}{\p t} =  -\frac{\p}{\p x_k}(\rho v_k) \qquad, 
\tag{12}  
\eal
\medskip  
Equation (20) is the  equation of  motion, where  $v_i$ is the $i^{th}\  (i = 1,2,3)$\\
 component of the velocity field, which is assumed to be a function of the\\
spatial coordinate $x_k$ as well as time  $t$.  Repeated sum notation is\\
 assumed.  As usual, $p$  is the pressure, $\rho$  is the mass density and $T$\\
is the temperature. The heat equation, Eq. (11) is written in terms of \\
the entropy per unit mass $s$, in order to explicitly show those terms that\\
instigate entropy production. The quantities $\si_{ik}$ and $q_k$  are the \\
viscosity stress tensor and the conductive heat flux density respectively. \\
They are given by the following constitutive relations:
\bal
& \si_{ik}= \et\, \, (\frac{\p v_i}{\p x_k} + \frac{\p v_k}{\p x_i}  -\frac23\, \de_{ik} \, \frac{\p v_t}{\p x_t})
+ \ze\, \, \de_{ik} \,  \frac{\p v_t}{\p x_t}
\tag{13}
\\[4.5ex]
&q_k= -\ka\ \frac{\p T}{\p x_k}\qquad, \tag{14}
\eal

 \clearpage

\vspace*{-1.0 in}

\centerline{-6-}

\vspace*{.65 in}
  
where $\et$ and $\ze$ are the viscosity coefficients, which are in general \\
positive functions of $p$  and $T$. The thermal conductivity is $\ka$, which \\
may in addition be a function of $|\nabla T|$. The system of equations given by\\
(10)-(12) is closed by the thermodynamic relations
\medskip
\[
p=\rho^2 \frac{\p U}{\p \rho} \eqno(15)
\]
\medskip
and
 \medskip
\[
T=  \frac{\p U}{\p s} \qquad, \eqno(16)
\]
\medskip 
where $U(\rho,s)$ is the internal energy per unit mass; $U(\rho,s)$ is assumed\\
to be a known function of $\rho$ and $s$.

\hspace{.4in}The Navier-Stokes equations, as given, are known to conserve the energy 
 \medskip
\[
\calh= \int (\frac12 \, \rho v^2 + \rho   U(\rho,s) ) \ d^3x \qquad,  \eqno(17)
\]
\medskip   
but produce entropy as a result of the terms of Eq. (11) involving $\si_{ik}$\\
 and $q_k$ ; i.e. by viscous dissipation and heat flux. Before presenting\\
 the symmetric bracket that produces these terms we review the Hamiltonian\\
structure for the Euler equations (i.e. $\si_{ik},\,  q_k \rightarrow 0$) as given in [17]\\
 (see also [2]).
 
 \hspace{.4in}The Hamiltonian in this case is the total energy functional of Eq. (17). \\
  The equations of motion, continuity and entropy are given by

   \clearpage

\vspace*{-1.25 in}

\centerline{-7-}

\vspace*{-.25 in}

\bal
&\frac{\p v_i}{\p t} =\{v_i,\calh\}  \tag{18}  
\\[1.5ex]
&\frac{\p \rho}{\p t} =  \{\rho,\calh\} 
\tag{19} 
\\[1.50ex]
&\frac{\p s}{\p t} =  \{s,\calh\} 
\tag{20}  
\eal
\medskip  
where the GPB, $\{\,,\,\}$, is given by
\bal
&\{F,G\}= - \int   \bigg(\frac{\de F}{\de \rho}\vec{\nabla}\cdot \frac{\de G}{\de \vec{v}}   +
 \frac{\de F}{\de \vec{v}}\cdot \vec\nabla  \frac{\de G}{\de \rho}  
 + \nn
\\[2.5ex]
&\frac{\de F}{\de \vec{v}}\cdot
\bigg[\frac{(\vec\nabla\times \vec{v})}{\rho} \times \frac{\de G}{\de \vec{v}}  \bigg]
+ \frac{\vec\nabla s}{\rho}\cdot \bigg[
\frac{\de F}{\de s} \frac{\de G}{\de \vec{v}} - \frac{\de F}{\de \vec{v}} \frac{\de G}{\de s} 
\bigg]\bigg) \, d^3x\qquad . 
\tag{21} 
\eal
Upon inserting the quantities shown on the right hand side of Eqs. (18)-(20),\\
 into Eq. (21) and performing the indicated operations one obtains, as noted,\\
the inviscid adiabatic limit of Eqs. (10)-(12).\\

 \hspace{.4in}The Casimirs for the bracket given by Eq. (21) are the total mass\\
$M=\int \rho \ d^3x$ and a generalized entropy functional $\cals_f =\int \rho f(s)\ d^3x$,\\
where $f$ is an arbitrary function of $s$. The latter quantity is added to\\ 
the energy [Eq. (17)] to produce the generalized free energy of Eq. (4):\\
$\calf=\calh + \cals_f$.

 \hspace{.4in}In order to obtain the dissipative terms, we introduce the following symmetric bracket:
\medskip
\bal
(F,G)&= \frac1{\la} \int   \bigg\{ \frac1\rho \frac{\de F}{\de v_i}\frac{\p}{\p  x_k}
 \bigg[\frac{\si_{ik}}{\rho}  \frac{\de G}{\de s}\bigg] + 
 \frac1\rho \frac{\de G}{\de v_i}\frac{\p}{\p  x_k}
 \bigg[\frac{\si_{ik}}{\rho}  \frac{\de F}{\de s}\bigg] 
  \nn
\\[1.5ex] 
&+ \frac{\si_{ik}}{T} \frac{\p v_i}{\p x_k} \bigg[\frac1{\rho^2}  \frac{\de F}{\de s} \frac{\de G}{\de s}\bigg]
+ T^2\ka \frac{\p }{\p x_k} \bigg[\frac1{\rho T}  \frac{\de F}{\de s} \bigg] \, 
\frac{\p }{\p x_k}  \bigg[\frac1{\rho T}  \frac{\de G}{\de s} \bigg]
\nn\\[1.5ex]
&+ T\ \La_{ikmn} \, \frac{\p }{\p x_m} \bigg[ \frac1{\rho}  \frac{\de F}{\de v_n} \bigg]
 \frac{\p }{\p x_k} \bigg[ \frac1{\rho}  \frac{\de G}{\de v_i} \bigg]
\bigg\} \, d^3x\qquad , 
\tag{23} 
\eal

 \clearpage

\vspace*{-1.0 in}

\centerline{-8-}

\vspace*{.65 in}

where
\medskip
\[
\La_{ikmn} = \eta\, (\de_{ni}\, \de_{mk}+ \de_{nk}\, \de_{mi}- \frac23\, \de_{ik}\, \de_{mn}) +\ze\ \de_{ik}\,\de_{mn}\,,
\eqno(24)
\]

from which we note that $\si_{ik}=\La_{ikmn}\, \p v_n/\p x_m$, and $\la$ is an arbirary \\ 
constant. In addition to symmetry this bracket possesses the following\\
 properties:
 
 \medskip

 \hspace{.4in}(a) There are degeneracies associated with the momentum functional\\
 $\vec{P}=\int \rho \vec{v} \, d^3x$ and energy functional $H$; i.e. $(\vec{P}, G)= (\calh, G)=0$ for all\\
 functionals $G$. 
 
 \medskip
 
\hspace{.4in} (b) For all functionals the bracket is definite with sign depending\\
upon $\la$. This is clear for the term that depends upon $\ka$  (recall $\ka > 0$), \\
but it is not immediately apparent for the remaining terms, so we rewrite\\
 the bracket as follows:
 \medskip
 \bal
 (F,G)&=   \frac1{\la} \int   \bigg\{ T\ \La_{ikmn} \,\bigg[ \frac{\p }{\p x_i} \bigg( \frac1{\rho} 
  \frac{\de F}{\de v_k} \bigg)- \frac1{\rho T} \frac{\p v_i}{\p x_k}   \frac{\de F}{\de s}\bigg]
 \nn\\
 &\times  \bigg[ \frac{\p }{\p x_m} \bigg( \frac1{\rho}  \frac{\de G}{\de v_n} \bigg)- \frac1{\rho T} \frac{\p v_m}{\p x_n}   \frac{\de G}{\de s} \bigg]
+ \ka T^2 \frac{\p }{\p x_k} \bigg[\frac1{\rho T}  \frac{\de F}{\de s} \bigg] 
\frac{\p }{\p x_k}  \bigg[\frac1{\rho T}  \frac{\de G}{\de s} 
\bigg\} \, d^3x\qquad .
\nn
\eal
 
 Definiteness arises from the fact that $\La_{ikmn} \ a_{ik} \, a_{mn} > 0$ for any $(a_{ik})$.\\
 An important ramification of definiteness occurs for the functional $\cals_f$. \\
 Definiteness in this case corresponds to an H-theorem, which is valid \\
 even though the function $f$ remains arbitrary.\\
 
 \medskip
 
 \hspace{.4in} (c) If we let $f = \la S$  upon inserting $\calf$ into Eq. (23) with $\vec{v}$,\\
$\rho$, and $s$ we obtain

\clearpage

\vspace*{-1.0 in}

\centerline{-9-}

\vspace*{.2 in}

\bal
&(v_j,\calf) =\frac1{\rho}\frac{\p}{\p x_k}\si_{jk} \tag{25}  
\\[2ex]
&(\rho,\calf)=0
\tag{26} 
\\[2ex]
& (s,\calf)=\frac{\si_{ik}}{\rho T}\frac{\p v_i}{\p x_k} + \frac1{\rho T} \frac{\p}{\p x_k} (\ka \frac{\p T}{\p x_k})
\tag{27}  
\eal
\medskip  
Equations (25)-(27) yield the dissipative terms of the Navier-Stokes\\
equations. Since $\cals$ is a Casimir, the Navier-Stokes equations are\\
given by
\bal
&\frac{\p v_j}{\p t} =\{\{v_j,\calf\}\}\nn
\\[2ex]
&\frac{\p \rho}{\p t} =  \{\{\rho,\calf\} \}
\nn
\\[2ex]
&\frac{\p s}{\p t} = \{ \{s,\calf\} \}
\nn
\eal
\medskip  
Observe that had we chosen a nonlinear $f$ Eqs. (25) and (27) would obtain \\
additional dependence upon $s$.

\hspace{.4in}In closing, we point out that for general systems, symmetry in transport \\
coefficients is related to bracket symmetry. For the purpose of illustration\\
we demonstrate this by replacing the scalar conductivity $\ka$ by a tensor\\
$\ka_{ij}$. Usually anisotropy arises because of. the presence of a magnetic \\
field $B$, as in the case of a crystal or conducting fluid. Here we\\
ignore the dependence of $\ka_{ij}$ on $B$, but evidently the formalism presented\\
here for the Navier-Stokes equations can be extended to magnetohydrodynamics\\
 with constitutive relations arising from small Larmor radius corrections [18].

 \clearpage

\vspace*{-1.0 in}

\centerline{-10-}

\vspace*{.65 in}

\hspace{.4in}If we replace the penultimate term of Eq. (23) by\\
\[
\int\bigg(T^2\  \ka_{ij}\  \frac{\p }{\p x_i}\bigg[\frac1{\rho T} \frac{\de F}{\de s} \bigg]
\frac{\p }{\p x_j}\bigg[\frac1{\rho T} \frac{\de G}{\de s} \bigg]
 \bigg) d^3x\qquad, 
 \eqno(28)
\]
then in order to maintain symmetry in the bracket it is necessary for\\
$\ka_{ij}=\ka_{ji}$. This corresponds to Onsager symmetry since here $\ka_{ij}(B)=$\\
$\ka_{ij}(-B)$. The contribution to the heat equation that is produced by \\
Eq. (28) is
\[
\frac1{\rho T} \frac{\p}{\p x_i}\  \ka_{ij}\  \frac{\p T}{\p x_j}\qquad.
\]

\underline{Acknowledgements}
\smallskip

\hspace{.4in}
I would like to acknowledge useful conversations with M. Grmela,\\
R.D.  Hazeltine, J.E. Marsden, R. Montgomery, T. Ratiu and W.B. Thompson. \\
This research was supported by DOE contracts DE-FG05-80ET-53088 and \\
DE-AT03-82ER-12097.

 }

\clearpage

\newgeometry{left=2.5cm, right=2.5cm, top=2.5cm, bottom=2.5cm}

 \underline{References}
 \bigskip

 \singlespace
 
 \begin{description}
 
\item 1. \ E.C.G. Sudarshan and N. Mukunda, \underline{Classical Dynamics: A Modern Perspective}\\
 (Wiley 1974).  Chapt. 9 deals with GPBs for ordinary differential equations.

\item 2. \ P.J. Morrison, "Poisson Brackets for Fluids and Plasmas", in Mathematical\\ 
Methods  in Hydrodynamics and Integrability in Dynamical Systems, eds.\\
M. Tabor and Y. Treve (AlP Conf. Proc. 88, 1982), 13-46.

\item 3. \ J. Marsden, A. Weinstein, T. Ratiu, R. Schmid and R. Spencer, ``Hamiltonian\\
 Systems with Symmetry, Coadjoint Orbits and Plasma Physics" in \underline{Modern}\\
  \underline{Developments in Analytical Mechanics}, (Proc. IUTAM-ISIMM, 1982).

\item 4. \ J.E. Marsden and' P.J. Morrison, ``Noncanonical Hamiltonian Field Theory and\\
 Reduced MHD", Contemporary Math. \underline{28} (1984). Am. Math. Soc.

\item 5. \ P .J. Morrison, ``Bracket Formulation for Irreversible Classical Fields", \\
Physics Letters \underline{100A}, 423 (1984).

\item 6. \ P.J. Morrison and R.D. Hazeltine, ``Hamiltonian Formulation of Reduced\\
 Magnetohydrodynamics" Physics of Fluids \underline{27}, 886 (1984).

\item 7. \ M. Grmela, talk presented at AMS workshop, Boulder, CO, July 1983,\\ 
``Particle and Bracket Formulations of Kinetic Equations", Contemporary\\
 Math. \underline{28} (1984). Am. Math. Soc.

\item 8.  \ A.N. Kaufman, ``Dissipative Hamiltonian Systems: A Unifying Principle", \\
Physics Letters \underline{100A}, 419 (1984).

\item 9.  \ P. Similon, private communication.

\item 10. \ M. Grmela, ``Bracket Formulation of Dissipative Fluid Mechanics Equations", \\
preprint (1984).

\item 11. \ W. Newcomb, appendix of I. Bernstein, Physical Review \underline{109}, (1958).

\item 12. \  M.D. Kruskal and C.R. Oberman, ``On the Stability of Plasma in Static\\
 Equilibrium", Physics of Fluids \underline{1}, 275 (1958).

\item 13.  \  T.K. Fowler, ``Lyapunov's Stability Criteria for Plasmas", J. Math. Phys.\\
\underline{4}, 559 (1963).

\item 14. \ M. Rosenbluth, "Topics in Microinstabilities", in \underline{Advanced Plasma Theory},\\
 (Academic 1964) 136.

\item 15. \  D.D. Holm, J .E. Marsden, T. Ratiu and A. Weinstein, "Nonlinear Stability\\
Conditions and A Priori Estimates for Barotropic Hydrodynamics", Physics\\
 Letters \underline{98A}, 15 (1983). (Also in preparation ``A Priori Estimates for Nonlinear Stability of Fluids and Plasmas".)

\end{description}
 
\clearpage

\begin{description}

\item 16. \  R.D. Hazeltine, D.D. Holm, J.E. Marsden and P.J. Morrison, ``Generalized \\
Poisson Brackets and Nonlinear Liapunov Stability - Appliction to Reduced \\
MHD, ``Center for Pure and Applied Math. Report PAM-2l4/NSRMHD, Berkeley, \\
CA (1984). Submitted to ICPP proceedings Lausanne (1984).
 
\item 17. \  P.J. Morrison and J.M. Greene, "Noncanonical Hamiltonian Density\\
 Formulation of Hydrodynamics and Ideal Magnetohydrodynamics", Phys. Rev. \\
 Lett. \underline{45}, 790 (1980).

\item 18. \ W.B. Thompson,  \underline{An Introduction to Plasma Physics}, (Pergamon 1962) 227.

\end{description}
\clearpage

 \end{flushleft}

 \clearpage
\begin{centering}
\begin{figure}[htb]
\hspace{-.35in}\includegraphics[scale=.8]{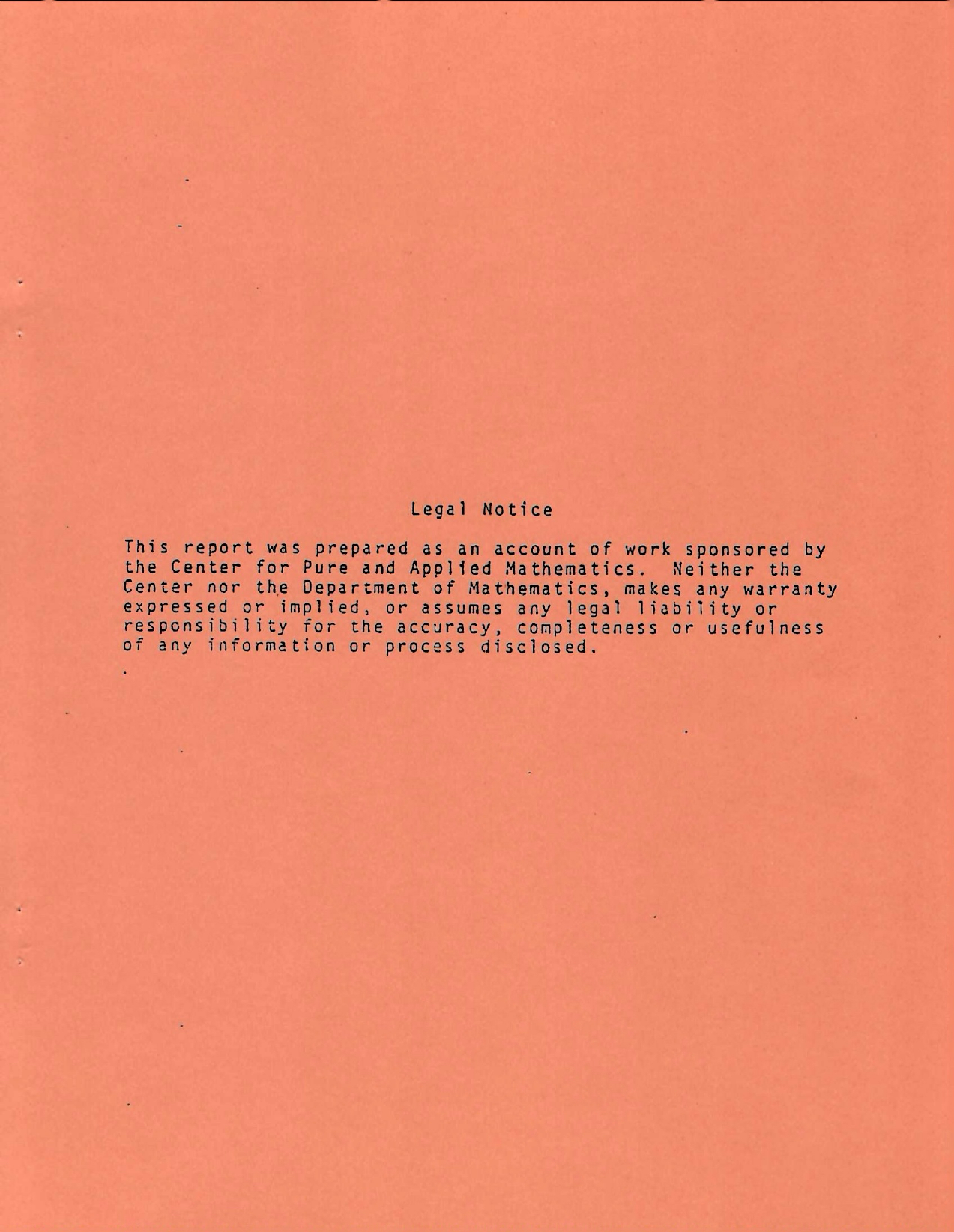}
\label{fig:tba}
\end{figure}
\end{centering}

\clearpage
\begin{centering}
\begin{figure}[htb]
\hspace{-.35in}\includegraphics[scale=.8]{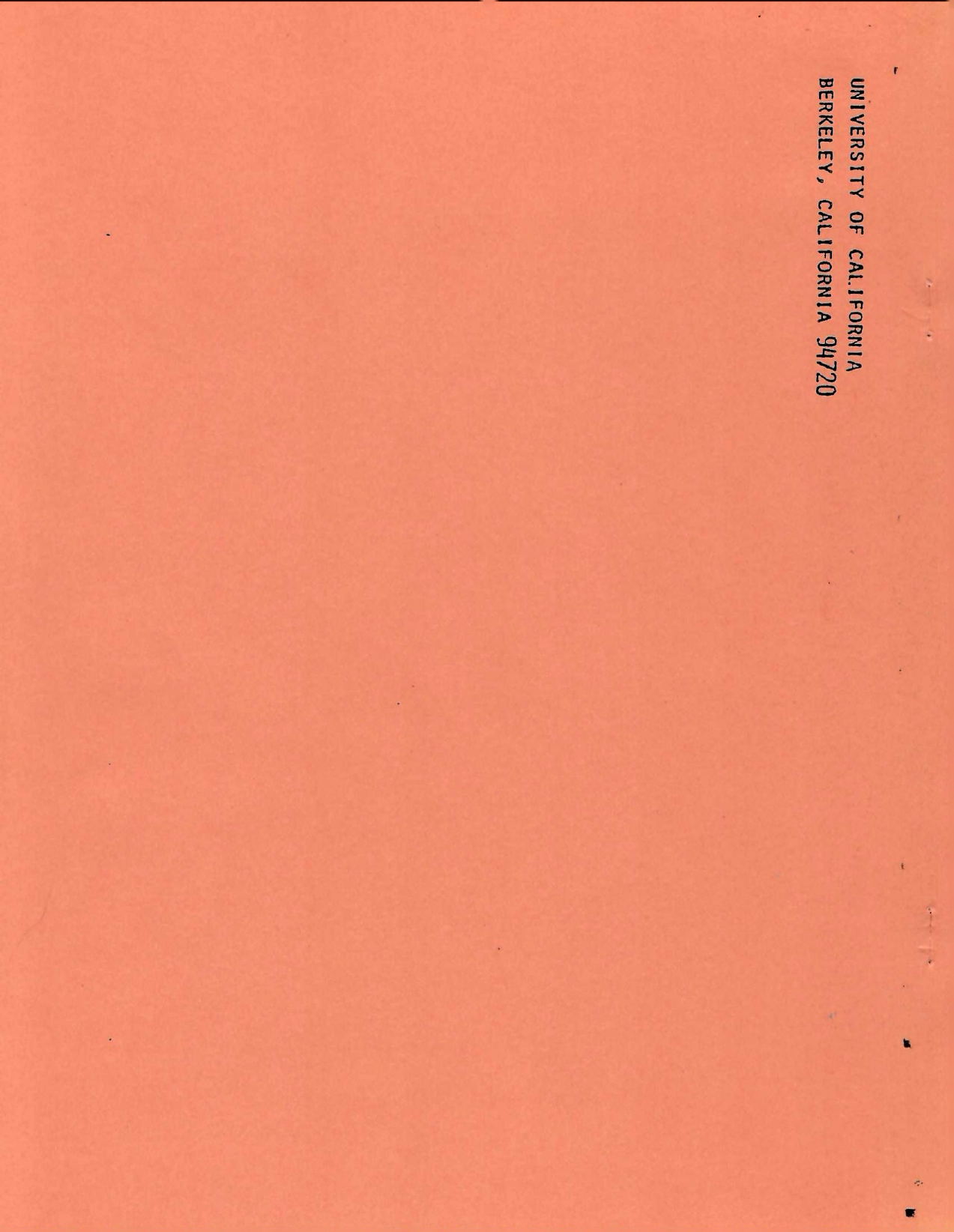}
\label{fig:tba2}
\end{figure}
\end{centering}

\clearpage

\section*{Errata }

\noindent * Page 1, 2 lines after eq.\ (2):  $\de\psi\rightarrow\de \psi^i$.

\noindent * Page 4, eq.\ (6) is misnumbered, resulting in two equations 6 and 7.

\noindent* Page 5, Line 2:  The ideas of degeneracy and definiteness in the context of brackets for dissipation  actually appeared earlier in   A. N. Kaufman and P. J. Morrison, ``Algebraic Structure of the Plasma Quasilinear Equations,''    Phys.~Lett {\bf 88A}, 405  (1982). 

\noindent* Page 8, Line 2 of (a): $H\rightarrow \calh$.
 
 \noindent* Page 12, Reference 16: Appliction $\rightarrow$ Application

\end{document}